\documentstyle[12pt]{article}
\begin{document}
\setlength{\unitlength}{1mm}
\newcommand{\I}{\mbox{\rm I} \hspace{-0.5em} \mbox{\rm I}\,}
\newcommand{\te}{\theta}
\newcommand{\be}{\begin{equation}}
\newcommand{\bee}{\begin{equation}}
\newcommand{\ee}{\end{equation}}
\newcommand{\ene}{\end{equation}}
\newcommand{\bea}{\begin{eqnarray}}
\newcommand{\ena}{\end{eqnarray}}
\newcommand{\tra}{\triangle\theta_}
\newcommand{\emt}{\varepsilon}
\newcommand{\ham}{\hat{{\cal H}}}
\newcommand{\inv}{\hat{{\cal I}}}

\begin{center}
{\bf MODELLING QUANTUM MECHANICS BY THE QUANTUMLIKE DESCRIPTION OF THE
ELECTRIC SIGNAL PROPAGATION IN TRANSMISSION LINES}
\end{center}

\begin{center}
{{\bf R.~Fedele$^1$,~M.A. Man'ko$^2$,~V.I.
Man'ko$^2$,~V.G. Vaccaro$^1$}}
\end{center}
\begin{center}
$^1${\it Dipartimento di Scienze Fisiche, Universit\'{a}
Federico II and INFN, Napoli,\\ Complesso Universitario
di Monte Sant Angelo, Via Cintia, I-80126 Napoli, Italy\\
e-mail:
renato.fedele@na.infn.it~~vittorio.vaccaro@na.infn.it }
\end{center}
\begin{center}
$^2${ \it P.N. Lebedev Physical Institute, Leninskii
Prospect 53,
 Moscow 119991, Russia\\
e-mail: manko@sci.lebedev.ru ~~~mmanko@sci.lebedev.ru}
\end{center}

\bigskip\bigskip\bigskip

\begin{center}
{\bf Abstract}
\end{center}

It is shown that the transmission line technology can be
suitably used for simulating quantum mechanics. Using
manageable and at the same time non-expensive technology,
several quantum mechanical problems can be simulated for
significant tutorial purposes.  The electric signal
envelope propagation through the line is governed by a
Schr\"{o}dinger-like equation for a complex function,
representing the low-frequency component of the signal,
In this preliminary analysis, we consider two classical
examples, i.e. the Frank-Condon principle and the
Ramsauer effect.

\section{Introduction}

There exist several purely classical systems whose
behaviour can be described with models which are fully
similar to the formal apparatus of quantum mechanics,
although their nature have nothing to do with the quantum
one. These models are usually referred to as "quantumlike
models" \cite{1,2}. Typically they are governed by a
Schr\"{o}dinger-like equation where $\hbar$ is replaced by
another physical parameter of the particular classical
system considered. In particular, quantumlike models have
been successfully proposed in the transport of
electromagnetic radiation beams in linear and nonlinear
regimes \cite{3,4} (see, for instance, the theory of
optical fibers \cite{5,6} and related subjects), in the
sound wave theory \cite{7}, in plasma physics \cite{8},
in the theories of transport and dynamics of
charged-particle beams \cite{9} and in dynamics of the
ocean waves \cite{10}. To describe the analytic signal
theory~\cite{11,12}, a quantumlike formalism has been
used; recently it was successfully employed to translate
new results of quantum mechanics into the theory of
analytic signal~\cite{13,14}. Quantumlike models received
also a big deal of attention in the literature to
describe the nonlinear electric signal propagation in
transmission lines in a number of papers. In particular,
this subject received a very important development in
literuture in connection with some experimental
investigations of modulational instability and soliton
formation in nonlinear transmission lines
\cite{15}.\newline Transmission line theory is also very
important for a number of scientific and technological
applications. In particular, it is widely used in
accelerator physics to model several parts of a given
accelerating machine \cite{16}. Also one should point out
that the transmission line can be considered as a
continuous set of short circuits with capacitance $C$,
inductance $L$, and resistance $R$. The separate short
circuit and two interacting short circuits were
considered in the quantum limit in~\cite{17}. The chain
of circuits was studied in the quantum domain
in~\cite{18}.\newline The quantumlike treatment of the
transmission line is the goal of our paper. For instance,
in this modelling operation, electric signals,
propagating through a transmission line, modelling a
piece of accelerating machines, may be analyzed to study
their characteristic impedances. It seems to be clear
that in the accelerator theory two "classical"
quantumlike problems are combined, namely, the analytic
signal theory and the propagation through the
transmission lines.

The aim of this work is pedagogical and consists in the
following steps: (i) to point out that the propagation of
an electric signal through a trasmission line can be
described with a quantumlike model in terms of a complex
wavefunction whose evolution is governed by a
Schr\"{o}dinger-like equation; (ii) to point out that our
results are relevant for simulating quantum effects with
manageable and non-expensive transmission line
technology.

The article is organized as follows. In the next section,
we briefly introduce the concept of a transmission line
and give the wave equation governing the propagation of a
signal as a current perturbation in space and time. The
effective {\it refractive index} of the line is also
introduced. In section 3, taking the {\it slowly-varying
amplitude approximation}, a Schr\"{o}dinger-like equation for
a complex function is derived from the wave equation. The
quantumlike formalism obtained in this way allows us in
section 4  to discuss the signal propagation in the cases
of some simple space profiles of the effective refractive
index that can simulate some tutorial problems of quantum
mechanics. For instance, we consider the case a quadratic
space-profile of the refrative index to show that the
quantum harmonic oscillator problem and the related
coherent states can be simulated by employing a
transmission line. In section 5, the perspective of how
to produce a simulation of more advanced quantum topics,
such as the Frank-Condon principle and the Ramsauer
effect, is drawn stressing the advantages that their
simulation can be obtained for tutorial purposes by using
a transmission line. Finally, conclusions and remarks are
presented in section 6.

\section{The wave equation of a linear dispersive transmission line}
Let us consider the usual representation of an
electromagnetic (e.m.) transmission line as a series of
R-L-C short circuits (series of moduli). Each modulus
exhibits inductance, capacitance and resistance per unity
length, $L'$, $C'$ and $R'$, respectively. Let us denote
with $x$ and $t$ the longitudinal space coordinate and
the time, respectively. Thus, it is easy to see that the
perturbation voltage and current signals, $\delta v(x,t)$
and $\delta i(x,t)$, appearing at the end of an arbitrary
modulus of the line, located at the longitudinal position
range $x, x+dx$, obey to the following coupled
short-circuit equations:
\begin{equation}
{\partial\delta v\over \partial x}~=~L'~{\partial\delta
i\over \partial t}~+~R'~\delta i~~~,
\label{voltage}
\end{equation}
\begin{equation}
{\partial\delta i\over \partial x}~=~C'~{\partial\delta
v\over \partial t}~~~.
\label{current}
\end{equation}
Combining (\ref{voltage}) and (\ref{current}), and solve
f.i. for $\delta i$ we get the following usual wave
equation for the transmission line:
\begin{equation}
{\partial^2\delta i\over \partial t^2}~-~{1\over
L'C'}~{\partial^2\delta i\over \partial x^2}~+~{R'\over
L'}~{\partial\delta i\over\partial t}~=~0~~~.
\label{delta-i-equation}
\end{equation}
This wave equation accounts for the dissipations along
the line of ohmic nature.\newline For the sake of
simplicity, let us assume that the ohmic dissipations are
negligible. Consequently, Eq. (\ref{delta-i-equation})
becomes:
\begin{equation}
{\partial^2\delta i\over \partial
t^2}~-~V^2~{\partial^2\delta i\over \partial x^2}~=~0~~~,
\label{delta-i-wave-equation}
\end{equation}
where we have introduced the phase velocity
\begin{equation}
V\equiv \sqrt{1\over L'C'}~~~.
\label{phase-velocity}
\end{equation}
When the parameters $L'$ and $C'$ of the line are
homogeneous, the phase velocity does not depend on the
coordinates, i.e.
\begin{equation}
V_0\equiv \sqrt{1\over L_{0}^{'}C_{0}^{'}}~~~,
\label{homogeneous-phase-velocity}
\end{equation}
where $L_{0}^{'}$ and $C_{0}^{'}$ are some unperturbed values.\newline In
order to take into account some very slow space and time modulations of
$L'$ and $C'$, let us assume:
\begin{equation}
L'=L_{0}^{'}f_{1}(x,t)~~~,
\label{L'}
\end{equation}
\begin{equation}
C'=C_{0}^{'}f_{2}(x,t)~~~,
\label{C'}
\end{equation}
where $f_{1}(x,t)$ and $f_{2}(x,t)$ are some specific
functions that account for the inhomogeneity profile in
space and time. Thus,
\begin{equation}
V^2~=~{1\over
L_{0}^{'}C_{0}^{'}f_{1}(x,t)f_{2}(x,t)}~\equiv~
{V_{0}^{2}\over N^2 (x,t)}~~~,
\label{phase-velocity-bis}
\end{equation}
where it is clear that $N(x,t)$ accounts for the
refractive index, say $n(x,t)$, of the {\it medium}.
\newline Note that: $V^2/c^2~=~V_{0}^{2}/c^2 N^2$ ($c$
being the light speed); thus
\begin{equation}
n(x,t)~=n_0 N(x,t)~~~,
\label{refractive-index}
\end{equation}
where $n_0$ is the {\it unperturbed} refractive index
(namely, the one of the homogeneous case). Consequently,
\begin{equation}
N(x,t)~={n(x,t) \over n_0}~~~,
\label{relative-refractive-index}
\end{equation}
is the relative refractive index.\newline By taking into account Eq.
(\ref{relative-refractive-index}), Eq. (\ref{delta-i-wave-equation})
becomes:
\begin{equation}
n^{2} (\eta,t)~{\partial^2\delta i\over \partial
t^2}~-~{\partial^2\delta i\over \partial
\eta^2}~=~0~~~,
\label{delta-i-wave-equation-bis}
\end{equation}
where $\eta =x/c$.
\section{A Schr\"{o}dinger-like equation for weakly-dispersive
transmission lines}

\subsection{The Telegraphist's equation}
In case of a time-independent refractive index, i.e.
$n=n(\eta)$, By taking a solution of
(\ref{delta-i-wave-equation-bis}) of the form
$$
\delta i = \left(\delta i\right)_0 \exp\left(-i\omega_0 t\right)~~~,
$$
we obtain the following Helmholtz-like equation (usually
referred as to "the telegraphist's equation"):
\begin{equation}
{\partial^2\delta i\over \partial
\eta^2}~+~c^2K^{2}(\eta)~\delta i~=~0~~~,
\label{telegraphist-equation}
\end{equation}
where $K^2 (\eta) \equiv \omega_{0}^{2}n^{2} (\eta)/c^2$.
A number of problems in linear transmission lines have
been described by means of eq.
(\ref{telegraphist-equation})\cite{16}.

\subsection{Slowly-varying amplitude approximation in
weakly-dispersive transmission lines} Now, in order to
describe the propagation of electric signal envelopes,
let us assume that $n(\eta,t)$ is weakly perturbed, i.e.
\begin{equation}
n(\eta,t)~\approx~n_0~+~\delta n (\eta,t)~~~,
\label{refractive-index-perturbation}
\end{equation}
where $|\delta n|<<n_0$. We look for a solution of
(\ref{delta-i-wave-equation-bis}) can be taken in the form:
\begin{equation}
\delta i(\eta,t)~=~\Phi (\eta,t)\exp \left(-i\omega t\right)~~~,
\label{delta-i}
\end{equation}
where $\Phi (\eta,t)$ is a very slow function of $t$
compared to the phase term variation, i.e.
\begin{equation}
\left|{\partial\Phi \over \partial s}\right|<<\omega |\Phi|~~~.
\label{slow-s-variation}
\end{equation}
The slow function $\Phi$ accounts for an amplitude
modulation of the signal, in such a way that $\delta i$
plays the role of a wave envelope (electric signal
envelope). Substituting (\ref{delta-i}) and
(\ref{refractive-index-perturbation}) in
(\ref{delta-i-wave-equation-bis}) and taking into account
the first-order quatities only, we get the following
Schr\"{o}dinger-like equation for the current perturbation
envelope:
\begin{equation}
i\omega{\partial\Phi\over\partial t}~=~-{1\over
2n_{0}^{2}}{\partial^2\Phi\over
\partial \eta^2}~-~\omega^2{\delta n\over n_0} \Phi~-
~{\omega^2\over 2}\Phi~~~.
\label{schroedinger-like-equation}
\end{equation}
Putting
\begin{equation}
\Phi (\eta,t)~=~\Psi (\eta,t) \exp\left(-i\omega t/2\right)~~~,
\label{Psi-function}
\end{equation}
we finally get the following equation
\begin{equation}
i{\partial\Psi\over\partial s}~=~-{1\over
2n_{0}^{2}}{\partial^2\Psi\over
\partial \tau^2}~-~{\delta n\over n_0} \Psi~~~,
\label{schroedinger-like-equation-1}
\end{equation}
where the dimensionless variable $s=\omega t$ and $\tau
=\omega \eta$ have been introduced. \newline
Note that (\ref{schroedinger-like-equation-1})
corresponds to the usual one-dimensional Schr\"{o}dinger
equation with $\hbar
=1$, and where $s$ and $\tau$ replace the time $t$ and
the space-coordinate $x$, respectively. Consequently, we realize that the
relative refractive index perturbation $-\delta n /n_0$ plays the role of
an effective potential and $n_{0}^{2}$ plays the role of an effective mass.
Thus, the (\ref{schroedinger-like-equation-1}) can be cast as
\begin{equation}
i{\partial\Psi\over\partial s}~=~-{1\over
2n_{0}^{2}}{\partial^2\Psi\over
\partial \tau^2}~+~U(\tau,s)\Psi~~~,
\label{schroedinger-like-equation-bis}
\end{equation}
where
\begin{equation}
U(x,s)~=~-~{\delta n \over n_0}~~~.
\label{potential}
\end{equation}
It is easy to see that (\ref{potential}) can be also cast
as:
\begin{equation}
U(\tau,s)~=~\sqrt{L_{0}^{'}C_{0}^{'}\over L'(\tau,s) C'(\tau,s)}~-~1~~~.
\label{potential-bis}
\end{equation}
For the sake of simplicity, in the following we fixe to
the unity the constant $n_0$. Note that, as in Quantum
Mechanics the wavefunction of an elementary particle is a
solution of the Schr\"{o}dinger equation, the complex
function $\Psi (\tau,t)$ involved in the
(\ref{schroedinger-like-equation-bis}) represents the
analog of the quantum wavefunction which is associated
with the electric signal envelope propagating through the
transmission line. In this paper we conventionally call
this complex function the {\it signal envelope
wavefunction} (SEW). Once the signal envelope
wavefunction is normalized, i.e.
\begin{equation}
\int_{-\infty}^{\infty}~|\Psi (\tau,s)|^2~d\tau~=1~~~,
\label{Psi-normalization}
\end{equation}
the quantity $|\Psi (\tau,s)|^2$ plays the role of
probability density to find the electric signal envelope
at the location $x=c\tau$ and at the time $t=s/c$. We may
syntetically call $|\Psi|^2$ the {\it probability density
associated with the electric signal envelope}.

\section{An examples of tutorial interest:
propagation through a line with a quadratic space-profile
refractive index} In this section we consider a special
case of effective refractive index that may be useful to
simulate some of the typical quantum problems of tutorial
interest. In fact, we examine the propagation of an
electric pulse through a transmission line with an
effective quadratic space-profile refractive index.

In case $\delta n$ is time-independent, an interesting
case to be considered is the one in which $-\delta n
/n_0$ is a parabolic function of $\tau$, namely ($n_0 =1$)
\bee
i {\partial \over \partial s} \Psi(\tau,s) =
- { 1 \over 2} { \partial^2 \over \partial \tau^2} \Psi(\tau,s)+
{1 \over 2} k~\tau^2 \Psi(\tau,s)~~~,
\label{6}
\ene
where we have assumed $-\delta n /n_0 \equiv k\tau^2
/2$, with $k>0$. It is easy to show that Eq. (\ref{6}) admits the
following orthonormal discrete modes for the SEW
\begin{eqnarray}
\Psi_{n}(\tau,s)  =   \frac{1}{\left[2 \pi~\sigma^{2}(s)~ 2^{2n}~(n!)^2
\right]^{1/4}}~H_{n}\left( \frac{\tau}{\sqrt{2} \sigma(s)}\right)
\nonumber\\
 {\times}  \exp\left[ - \frac{\tau^2}{4 \sigma^{2}(s)} + i
\frac{\tau^2}{2 \rho(s)} + i (1+2n) \phi(s) \right]~~~~\mbox{with}~~~
n=0,1,2,..~~~,
\label{13c}
\end{eqnarray}
where $H_{n}$ are the Hermite polynomials, $\sigma(s)$ obeys to the
following {\it envelope equation}
\bee
\sigma '' + k~\sigma - { 1 \over 4
\sigma^2}=0~~~,
\label{14c}
\ene
and
\bee
\frac{1}{\rho}  = \frac{\sigma '}{\sigma}~~~,
\label{15ca}
\ene
\bee
\phi ' = -{ 1 \over 4 \sigma^3} ~~~.
\label{16c}
\ene
where each prime denotes the derivative with respect to $s$. It is easy to
see that $\sigma(s)$, appearing in (\ref{13c})-(\ref{16c}), coincides with
the r.m.s. of the fundamental mode $\Psi_{0}(\tau,s)$; in general, an
arbitrary SEW $\Psi$ has a r.m.s. $\sigma (s)$ defined by:
\bee
\sigma(s) = \left[\int_{-\infty}^{+\infty}
\tau^2 \left| \Psi_{0}(\tau,s)\right|^2~d\tau \right]^{1/2}~~~.
\label{18c}
\ene
In addition, we can also define the expectation value for the transverse
linear momentum associated to an arbitrary SEW $\Psi(\tau,s)$
\bee
\sigma_{p}(s) = \left[\int_{-\infty}^{+\infty}
\left| { \partial \Psi_{0}(\tau,s) \over \partial \tau} \right|^2~
d\tau \right]^{1/2}~~~.
\label{19c}
\ene
where $\hat{p}=-i{\partial\over\partial\tau }$; in fact, $\tau$ and $p$
play the role of conjugate variables. \newline It is suitable to introduce
the following matrix
\bee
\hat{T}(s) \equiv \left( \begin{array}{cc} \sigma_{p}^2(s) &
- \sigma(s)~\sigma '(s) \\
- \sigma(s)~\sigma '(s) & \sigma^2(s) \end{array}\right)~~~,
\label{24c}
\ene
whose determinant is an invariant, namely
\bee
\sigma_{p}^2 \sigma^2 - \left( \sigma \sigma '
\right)^2 = {1\over 4} = \mbox{const.}~~~.
\label{25c}
\ene
It is easy to prove that
\bee
\sigma(s)\sigma '(s)= \int_{-\infty}^{+\infty}
\Psi_{0}^{*}(\tau,s) \left( { \tau \hat{p} + \hat{p} \tau \over 2}\right) \Psi_{0}(\tau,s)
~d\tau = \langle { \tau \hat{p} + \hat{p} \tau \over 2} \rangle ~~~.
\label{26c}
\ene
Consequently, from (\ref{25c}) follows that
\bee
\langle \tau^2 \rangle ~ \langle \hat{p}^2 \rangle -
\langle { \tau \hat{p} + \hat{p} \tau \over 2} \rangle^2 =
{1\over 4}~~~,
\label{27c}
\ene
which is formally identical to Robertson-Schr\"{o}dinger
uncertainty relation \cite{19}, \cite{20} for partial
cases when $~\langle \tau\rangle
=~\langle p\rangle =0$. Note that (\ref{25c}), or equivalently
(\ref{27c}), gives the usual form of the Heisenberg-like
uncertainty principle which is analogous to Heisenberg
uncertainty relation in quantum mechanics (again for
$~\langle
\tau\rangle
=~\langle p\rangle =0$)
\bee
\sigma_{p} \sigma \geq {1 \over 2}~~~.
\label{28c}
\ene
The equilibrium solution of (\ref{14c}) ($d^2
\sigma(s)/ds^2=0$), namely
\bee
\sigma_{0}^2 = { 1 \over 2 \sqrt{k} }~~~,
\label{32c}
\ene
implies that the set of Hermite-Gauss modes (\ref{13c}) reduces to the
hamiltonian eigenstates of the harmonic oscillator
\bee
\Psi_{n}^{0}(\tau,s) = { 1 \over [ 2 \pi \sigma_{0}^2 2^{2n} (n!)^2]^{1/4}}
\exp\left( - { \tau^2 \over 4 \sigma^2_{0}} + i (1+2n) \phi_{0}(s) \right)~
H_{n}\left( { \tau \over \sqrt{2} \sigma_{0} }\right)~~~,
\label{7}
\ene
where $n=0,1,2,....$,
\bee
\phi_{0}(s)=- \sqrt{k} {s \over 2} ~~~,
\label{8}
\ene
and the energy values ${\cal E}_{n}^{0}$, given by
averaging the Hamiltonian of the systen with the
wavefunction (\ref{7}), are the analog of the hamiltonian
eigenvalues of the quantum harmonic oscillator
\bee
{\cal E}_{n}^{0} = \left( n + { 1 \over 2} \right)
\sqrt{k}~~~.
\label{8ab}
\ene
In particular, for $n=0$ (\ref{7}), (\ref{8}) and (\ref{8ab}) give
the ground-like state
\bee
\Psi_{0}^{0}(\tau,s) = { 1 \over [ 2 \pi \sigma_{0}^2]^{1/4}}
\exp\left( - { \tau^2 \over 4 \sigma^2_{0}} + i  \phi_{0}(s) \right)~~~,
\label{8a}
\ene
which is purely Gaussian and the lowest energy reachable
by the electric signal envelope is ${\cal E}_{0}^{0} =
(1/ 2)
\sqrt{k}$. The means $\langle \tau\rangle $ and $\langle p\rangle $
are equal to zero at this state of the electric signal
envelope. In these conditions the uncertainty relation is
minimized as
\bee
\sigma_{0} ~ \sigma_{p0} = { 1\over 2}~~~.
\label{8ac}
\ene
Eq. (\ref{8ac}) holds also during the evolution of the electric signal,
because, in addition to (\ref{32c}), we have $\sigma_{p}(s)=\sigma_{p0}=
\mbox{const.}$.
In summary, we conclude that if we initially prepare the
SEW according to the matching conditions (\ref{32c}), its
evolution is ruled by a quantum-like behaviour in terms
of ground-like state which minimizes the uncertainty
relation and corresponds to the lowest accessible energy
$(1/ 2)\sqrt{k}$ of the electric signal envelope.

As it well known, SEW (\ref{8a}) belongs to the infinite
series of coherent state functions, labeled by a complex
number $\alpha =
\alpha_{1} + i \alpha_{2}$, and widely used in
quantum mechanics and quantum optics \cite{21,22}.

\section{Quantumlike Analogs of Frank--Condon and
Ramsauer--Twonsend Effects for Transmission Lines}

Let us discuss what physical consequences can be
extracted from the observation that the electric signal
in the transmission line can be associated with the
Schr\"{o}dinger-like equation. We got the result that
transmission line can be considered as a quantumlike
system; the distributed along the line conductance $C$
and inductance $L$, in principle can be considered as
inhomogeneous and time-dependent functions, i.e.,
$C=C(\tau,s)$ and $L=L(\tau,s)$. They account for the
effective potential-energy function $U=U(\tau,s)$. In
fact, in the quantumlike systems, such as the light ray
in optical fiber or electron beam in accelerator in the
framework of thermal-wave model, the refractive index
profile plays a role of the effective potential-energy
function.
\newline Let us consider now the interesting situation
in the presence of some filtering of the modes in the
transmission line. To this end, let us consider a
transmission line whose refractive index gives an
effective potential well $U$ wich has a rectangular
structure in domain of nonhomogenuity of the transmission
line. In this case, electrical signal in transmission
line is propagating along the line in complete analogy
with the electromagnetic wave in a waveguide. The
solutions to the Schr\"{o}dinger-like equations (modes of the
signal; in transmission line) can be treated as wave
functions which are reflected or transmitted by a
potential barier. There are effects involved in the above
propagation that we discuss qualitatively in the
following. We may consider this problem as an analog of
two known quantum subjects: The Ramsauer effect and the
Frank--Condon principle.

\subsection{Ramsauer effect}
In the case where there is non-homogeneity in the
transmission line, the potential-energy term has a
deformation, e.g., increasing (or decreasing) the depth
of the potential well for some length $L$. In the
wavelength of the signal function satisfies the condition
that $2L/\lambda=n$ where $n$ is even, the potential well
is transparent to the signal with wavelength $\lambda$.
It is just an analog of Ramsauer effect in which the
electron beam which scatters by atoms for some values of
energy (with corresponding de Broglie wavelength) has
smaller cross section for this process than for the other
wavelength when the ratio is odd. In the model
experiment, one can see that for some adapter a change in
frequency corresponds to different reactions of the
transmission line which just corresponds to the analog of
the Ramsauer effect. By this method, one can measure the
characteristics of the line (distributed inductance and
capacitance) that is equivalent to the impedance
measuring.

\subsection{Frank-Condon principle}
Another qualitative effect can be checked if one connects
two different pieces of the transmission line which
corresponds to connecting two potential bariers of
different depths. The problem of penetrating the modes of
the first piece and their transforming into the modes of
the second piece of the transmission line is equivalent
to calculating the Frank-Condon factor for electronic
transitions in polyatomic molecules. This factor is
overlap integral of the wave-like functions describing
the two different modes in the two pieces. $|C_{nm}|^2$
describes the portion of the signal energy of the $n$th
mode in the first part of the transmission line which
goes into $m$th mode of the second line. Again one can
pose the problem of maximality, e.g., of the
transformation of the fundamental mode energy into the
fundamental mode of the second pice of the line, having
in mind that distributed inductance and capacitance in
the second piece of the line are different from the first
one. In terms of introduced Wigner functions,  the
transformation coefficient $|C_{nm}|^2$ which give the
probability of transforming $n$th mode into $m$th mode
reads
\begin{equation}\label{Cnm}
|C_{nm}|^2=\int W_n(q,p)W_m(q,p)\,\frac{dq\,dp}{2\pi}.
\end{equation}
The coefficient $C_{nm}$ can be expressed in terms of overlap integral of
electric current modes $\psi_n(\tau)$, $\psi_m(\tau)$
$$C_{nm}=\int \psi_n^*(\tau)\psi_m(\tau)\,d\tau,$$
where the mode $\psi_n(\tau)$ is the mode in the first
piece of the transmission line and the mode
$\psi_m(\tau)$ corresponds to the second piece of the
transmission line. If one can vary in time the inductance
and capacitance of the transmission line, the
Frank-Condon factor describes the parametric excitation
of the modes in the transmission line. For example, if
the rectangular profile of `refractive index' is modeled
by the parabolic profile with the varying frequency, the
electric current propagation in the line is described by
the oscillator with time-dependent frequency used for
analysis of the thermal wave model~\cite{23}. We can
adopt the results of the analysis and apply them to the
transmission line. Thus, we have for the fundamental mode
the Gaussian solution~\cite{23}
\begin{equation}
\psi_0(\tau,t)=\pi^{-1/4}\varepsilon^{-1/2}(t)\exp\left(\frac{i\dot\varepsilon
(t)\tau^2}{2\varepsilon(t)}\right),
\end{equation}
where the function of time $\varepsilon(t)$ satisfies the equation
\begin{equation}
\ddot\varepsilon(t)+\omega^2(t)\varepsilon(t)=0
\end{equation}
and the initial conditions
$$\varepsilon(0)=1\qquad \dot\varepsilon(0)=i$$
for units chosen in such a manner that $\omega(0)=1$. The
frequency $\omega(t)$ corresponds to parabolic shape of
the `refractive index' $\omega^2(t)\tau^2/2$ of the
transmission line. In the case of instantaneous change of
the line parameters in time, one has excitation of the
line modes given by the Frank-Condon factor. If the
change is not instantaneous one can get the modes in the
form~\cite{23}
\begin{equation}
\psi_n(\tau,t)=\psi_o(\tau,t)\left(\frac{\varepsilon (t)}{\varepsilon^*(t)}\right)
^{n/2}\frac{1}{\sqrt{2^nn!}}\,H_n\left(\frac{\tau}{|\varepsilon(t)|}\right),
\end{equation}
where $H_n$  is Hermitte polynomial. the Frank-Condon factor can be
calculated in the explicit form for the model under consideration. Thus,
the coefficients $C_{nm}$ are expressed in terms of two-dimensional Hermite
polynomials. The Frank-Condon factors qualitatively can be evaluated by a
geometric method. As it is known, the maximally excited is the mode for
which refractive index curve after the change of the line parameters is
intersecting the perpendicular taken from the rest point of the initial
refractive index curve. Thus the modes are connected which have common rest
points on the plot of two curves. One curve is the initial refractive index
and the other one is the final refractive index.

\section{Conclusions and perspectives}
We have shown that the electrical current signal in the
transmission line obeys to the Schr\"{o}dinger-like equation.
We have extracted some physical consequences from the
fact that the transmission line is the quantumlike
system. In our preliminary analysis, we have considered
two physical situation of a tramsmission line that can be
thought as the quantum analog of known quantum phenomena.
The first physical situation involves the analog of the
Ramsauer effect which explains the behaviour of the line
signal if one considers two pieces of the line connected
by some adapter. In this case, one can filter the modes
in order to increase the transmission of some modes and
suppress some others. The other physical situation
involves the quantum analog of the parametric excitation
of the modes in the line which corresponds to quantum
transitions in polyatomic molecules described by the
Frank-Condon factor. Remarkably, this approach seems to
be very helpful and promising for providing a very simple
method of simulating a number of quantum-mechanical
effects, using both manageable and non-expensive
technology.

\end{document}